\documentstyle[12pt]{article}

\newcommand{\bb}{\begin{equation}}
\newcommand{\ee}{\end{equation}}
\newcommand{\bqn}{\begin{eqnarray}}
\newcommand{\eqn}{\end{eqnarray}}

\begin{document}
\begin{titlepage}

\begin{flushright}

ULB-TH-99/08 \\
LPT-ORSAY 99-57

\end{flushright}

\begin{center}
{\Large {\bf  Semi-Invariant 
Terms for Gauged Non-Linear $\sigma$-Models}}

\end{center}
\vfill

\begin{center}
{\large
Marc Henneaux$^{a,b}$ and Andr\'e Wilch$^a$
\footnote{henneaux@ulb.ac.be, awilch@ulb.ac.be}}
\end{center}
\vfill

\begin{center}{\sl
$^a$ Physique Th\'eorique et Math\'ematique, Universit\'e Libre de
Bruxelles,\\
Campus Plaine C.P. 231, B--1050 Bruxelles, Belgium\\[1.5ex]

$^b$ Centro de Estudios Cient\'\i ficos de Santiago,\\
Casilla 16443, Santiago 9, Chile\\[1.5ex]

}\end{center}

\vfill

\begin{abstract}
We determine all the terms that are gauge-invariant
up to a total spacetime derivative
(``semi-invariant terms") for gauged non-linear sigma
models.
Assuming that the isotropy subgroup $H$
of the gauge group is compact or semi-simple,
we show that (non-trivial) such terms exist
only in odd dimensions and are equivalent to the 
familiar Chern-Simons terms for 
the subgroup $H$.
Various applications are mentioned, including one to the
gauging of the Wess-Zumino-Witten terms in even 
spacetime dimensions.
Our approach is based on the analysis of the descent equation associated with
semi-invariant terms.

\end{abstract}
\vfill
\end{titlepage}

\section{Introduction}
\setcounter{equation}{0}

The Lagrangian of a dynamical system invariant under a
given symmetry is invariant itself
up to a total spacetime derivative, $\delta_\epsilon {\cal L} =
\partial_\mu k^\mu$.  In some cases, one can
redefine the Lagrangian without modifying its Euler-Lagrange
derivatives, ${\cal L} \rightarrow {\cal L}' = 
{\cal L} + \partial_\mu l^\mu$, so that the new Lagrangian is strictly
invariant, $\delta_\epsilon {\cal L}' = 0$.  In other
cases, however, this is impossible.  A well-known example
is the Chern-Simons term for Yang-Mills theory in
odd spacetime dimensions \cite{DeJaTe}, which is invariant only up
to an unremovable total derivative.  Terms for which
$\delta_\epsilon m = \partial_\mu k^\mu$  but $m \not=
m' - \partial_\mu m^\mu$ with $\delta_\epsilon m' = 0$ are
sometimes called (non trivial)
``semi-invariant terms" and we shall
adopt here this convenient terminology.

Given a theory, it is important to know all the terms that can be
added to the action without destroying the assumed symmetries.
While the identication of the manifestly invariant terms
is usually rather straightforward, it is in general more 
difficult to determine systematically the semi-invariant ones.
The purpose of this letter is to solve this question 
for gauged non-linear
sigma models.
As it is well known, these models occur repeatedly 
in supergravity \cite{Ca,Hull,PerSez}.  

Our work is motivated by the
beautiful paper \cite{deHuRo}, where a method for constructing
a new class of semi-invariant terms in even spacetime dimensions, with
topological connotation, is described.  As pointed out there, a necessary
condition for these terms to exist is that the gauge group
be non
semi-simple.  The question was further studied
in \cite{HeWi}, where it was proved that if 
the isotropy subgroup $H$ of the gauge group reduces to the identity,
then, no matter what
the gauge group is, the terms of \cite{deHuRo}
differ from strictly gauge-invariant terms 
by a total derivative and thus are not truly
semi-invariant.  We show here that this conclusion
remains valid whenever $H$ is compact or
semi-simple.
In fact,    
the only available non-trivial semi-invariant terms are then the Chern-Simons
terms based on the subgroup $H$, but these exist only in odd
dimensions.  More precisely, any semi-invariant   
term is proportional to a Chern-Simons term up to terms that are
strictly gauge-invariant and up to a total derivative.
Thus, under the assumption that
$H$ is compact or semi-simple, non-trivial semi-invariant terms 
are exhausted by the 
familiar Chern-Simons family.

Our approach is based on the analysis of the BRST
descent equation associated with
semi-invariant terms.

\section{Description of $\sigma$-Models}
\setcounter{equation}{0}

We denote the gauge
group by $G$, its Lie algebra by ${\cal G}$ and
the Yang-Mills connection by $A^A_\mu$. 
The Lie group $G$ may be non-compact since non-compact groups
are typically the rule in supergravity.  In addition
to the sigma-model variables,
we allow for the presence of
``matter" fields $y^i$ transforming
in some linear representation of $G$.

The gauge group $G$ may not act transitively on the manifold
$M$ of the non-linear sigma model.  For instance, $M$ 
could be a homogeneous space $G'/H'$ invariant under a bigger
group $G'$ of which $G$ is only a proper subgroup. 
In that case, we assume the existence of a slice
through every point of $M$, so that $M$ 
splits into $G$-orbits, which are 
smooth submanifolds \cite{slices}.  
The $G$-orbits are - by definition - homogeneous 
spaces for $G$ and thus of the form $G/H$ where the isotropy subgroup $H$
depends in general on the orbits.
In a stratum where the isometry subgroup is constant, one
may locally introduce coordinates 
that parametrize the points in the coset manifold $G/H$
and coordinates $\psi^\Delta$ that are constant along the orbits\footnote{ 
How to patch strata together
may presumably be achieved along the lines of \cite{Hull} combined
with the results concerning linear
representations, but we have
not investigated this question here.}.  
These latter transform in the trivial
linear representation of $G$ and
can simply be included among the ``matter fields"
$y^i$'s.  They are gauge-invariant while the coordinates
along the orbits could be gauged away by fixing the gauge
(as we shall see,
the $\psi^\Delta$'s contribute to the BRST cohomology but not the
coordinates along the orbits).

Thus, provided one allows for extra matter
fields and includes the coordinates $\psi^\Delta$
among them, one can assume that $G$ (the gauge group) acts transitively
on the (smooth) manifold of the non-linear model, which
has the form $G/H$.
This will be done from now on.
The subgroup $H$ is
assumed to be compact or semi-simple.  

First, we recall a few properties of coset spaces.  For that
matter, it is convenient to split the generators $T_A$ of ${\cal G}$ 
into two sets, the generators of
the Lie algebra of $H$ (${\cal H}$)  and the remaining ones,
\begin{equation}
T_A=\{T_{\alpha},T_a\},
\label{split_of_generators}
\end{equation}
where capital latin indices refer to ${\cal G}$, small greek indices
to ${\cal H}$ and small latin
indices to the chosen supplementary subspace ${\cal K}$. The commutators
\begin{equation}
[T_A,T_B]=f^C_{AB}T_C, \ \
[T_{\alpha},T_{\beta}]=f^{\gamma}_{\alpha\beta}T_{\gamma}
\end{equation}
define the structure constants of ${\cal G}$ and ${\cal H}$ respectively.
Because $H$ is semi-simple or compact, the coset space is reductive.
This means that one can choose the supplementary generators $T_a$
so that the commutators $[T_a, T_{\alpha}]$ involve only the $T_b$'s,
\begin{equation}
[T_a,T_{\beta}]=f^c_{a\beta}T_c.
\end{equation}
Thus, the structure constants $f^{\gamma}_{a\beta}$ vanish
and the generators $T_a$ define a representation of $H$.

Coset representatives may 
be taken to be of
the form
  \begin{equation}
  U(\xi)=\exp\{\xi^aT_a\},
  \label{cosetrep}
  \end{equation}
with some real fields $\xi^a$, the number of which equals the dimension
of $G/H$.  
The $\xi^a$ provide local coordinates in the vicinity of
$\xi^a = 0$.  We shall in a first approach restrict the sigma-model
fields to be in a star-shaped neighbourhood of $\xi^a=0$.
How to deal with global features (which are easily incorporated)
is discussed in the conclusions.

The group $G$ acts on the left as
\begin{equation}
gU(\xi)=U(\xi')h(\xi,g)
\label{groupleft}
\end{equation}
with $g\in G$ and $h\in H$ (note that right action conventions
were adopted in \cite{HeWi}). The infinitesimal transformation
property of the local parameters
$\xi^a$ is derived through the parametrization
\begin{equation}
g=\exp\{-\epsilon^AT_A\}, \ \
h(\xi,g)=\exp\{-u^{\alpha}(\xi,\epsilon)T_{\alpha}\}
\end{equation}
leading to
\begin{equation}
(1-\epsilon^AT_A)U(\xi)=
(U(\xi)+\partial_aU(\xi)\delta\xi^a)(1-u^{\alpha}(\xi,\epsilon)T_{\alpha}).
\end{equation}
Setting
\bb
U^{-1} \partial_a U = \mu_a^A(\xi) T_A
\label{useful1}
\ee
and 
\bb
U^{-1} T_A U = k_A^B(\xi) T_B
\label{useful2}
\ee
one gets
\begin{equation}
\delta\xi^a=
\Omega^a_A(\xi)\epsilon^A,
\end{equation}
with $\Omega^a_A(\xi)$ defined through 
\bb
\Omega^a_A \mu_a^b = - k_A^b
\label{useful3}
\ee
[The matrix $\mu_a^b(\xi)$ is invertible in a vicinity of 
$\xi^a =0$ since one has 
$\mu_a^b(\xi) = \delta^b_a + O(\xi)$ from 
(\ref{cosetrep}) and (\ref{useful1}).
Note also $\mu_a^\alpha = O(\xi)$ and $k_A^B(\xi) = \delta_A^B +
O(\xi)$.  The matrices $k_A^B(\xi)$ are in fact matrices of
the adjoint representation of $G$ and thus clearly invertible].
The $\Omega_A=\Omega^a_A\partial_{a}$ are the Killing vectors
of the coset manifold.  They obey
the relation $[\Omega_A,\Omega_B]=f^C_{AB}\Omega_C$,
or in coordinates
\begin{equation}
\Omega^b_A\partial_b\Omega^a_B-\Omega^b_B\partial_b\Omega^a_A=\Omega^a_Df^D_{AB}
\end{equation}
as well as $\Omega_a^b = - \delta_a^b + O(\xi)$ and
$\Omega_\alpha^b = O(\xi)$.
Similarly, one gets
\bb
u^\alpha = \Omega_A^\alpha \epsilon^A
\label{useful4}
\ee
with
\bb
\Omega_A^\alpha(\xi) = k_A^\alpha + \mu_a^\alpha \Omega_A^a.
\label{useful5}
\ee
For later purposes, we observe that $\Omega_\beta^\alpha =
\delta_\beta^\alpha + O(\xi)$ and $\Omega_b^\alpha = O(\xi)$.
It follows that the matrix $\Omega_B^A$ is invertible (in the vicinity of
the identity), so that one can express the gauge parameters
$\epsilon^A$ in terms of the field variations $\delta \xi^a$ and the
$u^\alpha$'s.

The gauge transformations of the gauged non-linear $\sigma$ model $G/H$
(with $G$ the gauge group) read
\begin{equation}
\delta_\epsilon A^A_{\mu} = {\cal D}^{(A)}_{\mu} \epsilon^A, \; \;
\delta_\epsilon \xi^a = \Omega^a_A \epsilon^A
\label{gaugetransfsigma}
\end{equation}
where ${\cal D}^{(A)}_{\mu}$ is the covariant derivative with
connection $A^A_\mu$, ${\cal D}^{(A)}_{\mu} \epsilon^A =
\partial_{\mu}\epsilon^A+f^A_{BC}A^B_{\mu}\epsilon^C$.
The question is to find the most general gauge invariant 
function of the fields and their derivatives up to a total divergence.

\section{BRST Differential}
\setcounter{equation}{0}

If ${\cal L}$ is invariant under gauge transformations up
to a total derivative,
\bb
\delta_\epsilon {\cal L} =
\partial_\mu k^\mu
\ee
then $k^\mu$ itself is not arbitrary but subject to
some definite conditions obtained by evaluating the commutators of the
second variations
of ${\cal L}$.

It is convenient to analyse this question in terms of the BRST
differential taken here to act from the left and
explicitly defined through
\begin{eqnarray}
\gamma A^A_{\mu} &=& {\cal D}^{(A)}_{\mu}C^A, \\  
\gamma \xi^a &=& \Omega^a_AC^A,  \\ 
\gamma y^i &=& - Y_{Aj}^i y^j C^A, \\
\gamma C^A &=& - \frac{1}{2}f^A_{BC}C^BC^C.
\end{eqnarray}
where the matrices $Y_A$ are the generators of the representation
of the $y$'s, $[Y_A, Y_B] = f^A_{BC} Y_C$,  and where $C^A$ are the ghosts.
The BRST differential acts also on the so-called antifields.  Since we are 
interested only in terms that are off-shell gauge invariant (up
to a total divergence), we shall, however, not include them.
Comments on more general deformations are briefly given at the end.

Local forms are exterior forms on spacetime which depend on the
fields, the ghosts and their derivatives.  A density ${\cal L}$ 
that is gauge-invariant
up to a total divergence defines, in dual terms, a local $n$-form
with ghost number zero that is $\gamma$-closed modulo $d$,
\begin{equation}
\gamma a + db = 0.
\label{cocycle}
\end{equation}
A redefinition of ${\cal L}$ by a total divergence is clearly equivalent
to
\begin{equation}
a \rightarrow a + \gamma c + de
\end{equation}
since there is no form in negative ghost number in
the absence of antifields ($c \equiv 0$). 
The question under study is thus equivalent to determining
whether elements in the cohomology $H^0(\gamma \vert d)$ have
a representative that is strictly $\gamma$-exact.

The advantage of reformulating the question in these
cohomological terms is that there are well-developed techniques
for handling the BRST cohomology $H^0(\gamma \vert d)$.  The idea is that 
the equation (\ref{cocycle}) implies $\gamma b + de = 0$ for some
$(n-2)$-form $e$ of ghost number two.  By continuing in the
same way, one gets a chain of equations $\gamma a + db = 0$,
$\gamma b + de = 0$, $\gamma e + df = 0$ (the ``descent"), the last 
two of
which read $\gamma m + dn  =0$ and $\gamma n = 0$.  If
$n$ is trivial in $H(\gamma)$, $n = \gamma u$, then one can
absorb it in redefinitions  ($m \rightarrow  m'=  m  - du$
and $n \rightarrow n - \gamma u = 0$)
and shorten the descent by one
step, the last equation being now $\gamma m' = 0$.

In a first stage, one thus
determines all non trivial solutions of $\gamma n =0$, without restriction
on the ghost number.  In a second stage, one determines which
among these solutions can be lifted all the way back up to
form-degree $n$.  Both problems have been studied in the literature.

\section{The $\gamma$-cohomology}
\setcounter{equation}{0}

Our strategy for computing the $\gamma$-cohomology is to perform
an appropriate change of variables in the ``jet-space" coordinatized
by the fields, the ghosts and their derivatives.  This change
of variables is adapted to the symmetry and combines the
analysis of \cite{all} for coset models with ungauged symmetry with the
change of variables of \cite{HeWi} for the gauged principal
models ($H$ reduced to the identity).

The starting point is the observation that if one decomposes
the Lie algebra-valued form $ U(\xi)^{-1}(d+A)U(\xi)$ in the
basis $(T_a, T_\alpha)$,
  \begin{equation}
  U(\xi)^{-1}(d+A)U(\xi)=p^a(\xi,A)T_a+v^{\alpha}(\xi,A)T_{\alpha},
  \label{central}
  \end{equation}
one generates a quantity $p^a_{\mu}$ that is covariant under the action of
$H$, and a field $v^{\alpha}_{\mu}$ that behaves as an $H$-gauge connection,
  \begin{eqnarray}
  \gamma p^a_{\mu}&=& - f^a_{\gamma b}p^b_{\mu}D^{\gamma} \\
  \gamma v^{\alpha}_{\mu}&=&f^{\alpha}_{\beta\gamma}
  v^{\beta}_{\mu}D^{\gamma}+
  \partial_{\mu}D^{\alpha}
  \end{eqnarray}
The ghosts $D^{\alpha}$ are identical to the 
$u^{\alpha}(\xi,\epsilon)$ that
occured in the parametrization of $h(\xi,g)$, 
with the gauge parameters $\epsilon^A$ being replaced
by the ghosts $C^A$,
  \begin{equation}
  D^{\alpha}(\xi)=u^{\alpha}(\xi,C)=\Omega^{\alpha}_B(\xi)C^B.
  \label{parametrization_1}
  \end{equation} 
The nilpotence of $\gamma$ requires the $D^{\alpha}$ to transform as
  \begin{equation}
  \gamma D^{\alpha}= -
  \frac{1}{2}f^{\alpha}_{\beta\gamma}D^{\beta}D^{\gamma}
  \label{main_condition}
  \end{equation}
We complete the redefinition of the ghosts by introducing 
the abelian ghosts $D^a$, which are the BRST variations
of the coset space coordinates $\xi ^a$,
  \begin{equation}
  D^a=\gamma\xi^a=\Omega^a_B(\xi)C^B, \ \ \gamma D^a=0
  \end{equation}
This redefinition of the ghosts is clearly invertible in the vicinity
of the identity, since the matrix $\Omega^A_B(\xi)$ is invertible.
Thus, one can trade the $C^A$ for the $D^A$,
  \begin{equation}
  D^A=\Omega^A_B(\xi)C^B, 
  \ \ C^A=\omega^A_B(\xi)D^B.
  \end{equation}
The $H$-connection $v$ can be used to
construct a $v$-covariant derivative of $p$,
  \begin{equation}
  {\cal D}^{(v)}_{\mu}p_{\nu}
  =\partial_{\mu}p_{\nu}+[v_{\mu},p_{\nu}], \ \ 
  \gamma{\cal D}^{(v)}_{\mu}p_{\nu}
  =[{\cal D}^{(v)}_{\mu}p_{\nu},D]
  \end{equation}
The suffix $v$ has been introduced to distinguish
${\cal D}^{(v)}_{\mu}$ from 
${\cal D}^{(A)}_{\mu}$.

The crucial observation now is that one can use (\ref{central})
to express the Yang-Mills connection $A^A_\mu$ in terms of
the variables $\xi^a$, $\partial_\mu \xi^a$,
$p^a_\mu$ and $v^\alpha_\mu$ (in the vicinity of the identity).
Indeed, the matrix multiplying $A^A_\mu$ in (\ref{central})
is the matrix $k_A^B$ of (\ref{useful2}), which is invertible.
The relation $A^A_\mu = A^A_\mu(\xi^a, \partial_\mu \xi^a,
p^a_\mu, v^\alpha_\mu)$ 
may clearly be prolonged to the derivatives of $A^A_{\mu}$, which thus may be
expressed in an invertible way through the 
ordinary derivatives of $\xi^a$ and $v^{\alpha}_{\mu}$
as well as the $v$-covariant derivatives of $p^a_{\mu}$. 
Furthermore, the relation between the
original ghosts $C^A$ and the redefined ones, $C^A=\omega^A_BD^B$, 
may also be prolonged to
the subsequent derivatives.  

In the same way, all the linearly
transforming fields $y^i$  may be combined with the coset
representative $U(\xi)$ written in the adequate  representation,
to form $H$-covariant fields together with their
$v$-covariant derivatives.  If for instance $y^i$ transforms in
the representation ${\bf D}$ of
$G$ generated by $(Y_A)^i_j$, $y'={\bf D}(g)y$,
then the quantity $\tilde{y}={\bf D}(U^{-1}(\xi))y$
transforms only under $H$,
$\tilde{y}'={\bf D}(h)\tilde{y}$, with $h$ given by (\ref{groupleft}). 
In terms of the BRST differential, $\gamma \tilde{y}^i
= - Y^i_{\alpha j} \tilde{y}^j D^\alpha$.
The $v$-covariant derivative
${\cal D}^{(v)}\tilde{y}$ shares the same transformation property.

Thus, one may take as
redefined set of jet-space coordinates the
local coset coordinates $\xi^a$ and the new ghosts $D^A$,
together with their ordinary derivatives; the
$H$-gauge connection $v^{\alpha}_{\mu}$ with its
ordinary derivatives; and the $H$-covariant quantities
$p^a_{\mu}$ and $\tilde{y}^i$  with their $v$-covariant derivatives.

The variables $\xi^a$ and $D^a$ form BRST contractible pairs
since $\gamma \xi^a = D^a$ and $\gamma D^a = 0$.  The same is
true for their derivatives.  Thus the BRST cohomology
involves only the $H$-connection $v^{\alpha}_{\mu}$ and the
the ghosts $D^\alpha$, as well as the
$H$-covariant objects $p^a_{\mu}$ and $\tilde{y}^i$, together
with their $v$-covariant derivatives.  The problem of computing 
$H(\gamma)$ for the Yang-Mills $+$ gauged non-linear sigma model $G/H$
with gauge group $G$ is thus reduced to the problem of 
computing $H(\gamma)$ for a Yang-Mills model with gauge group
$H$ and fields transforming in some definite
linear representation of $H$.  This is a well-known problem
whose solution may be found in \cite{DVTV,BDK,DVHTV}. 
Let $[\tilde{\chi}]_c$
denote the $H$-covariant fields $p^a_{\mu}$, $\tilde{y}^i$
and $F^{(v) \alpha}_{\mu \nu}$
and their successive $v$-covariant derivatives.  Then,
any $\gamma$-cocycle takes the form 
\begin{equation}
  \gamma a=0 \Rightarrow a=P^I_{inv}([\tilde{\chi}]_c)
\omega^{Lie}_I(D) + \gamma b
  \label{strict_inv_local}
  \end{equation}
where $P^I_{inv}$ is an invariant polynomial 
and $\omega^{Lie}_I(D)$ stands
for a basis of the Lie  algebra cohomology of $H$.
Since $H$ is compact or semi-simple, this cohomology is
generated by the ``primitive elements" which are the ghosts
of the abelian factors as well as $trD^3$ and traces of 
appropriate higher
powers of $D$ for the
non-abelian factors (see \cite{Greub} for more information).

Two extreme cases should be mentioned.
In the principal case ($H = \{e\}$), 
the variables $v^\alpha_\mu$ are absent and the
$p^a_\mu$ reduce to the $\tilde{I}^a_\mu$ of \cite{HeWi}.
At the opposite end, when $H = G$ ($G/G$-model), $U(\xi)= I$
(so that $\xi^a= \partial \xi^a = \cdots = 0$), the
variables $p^a_\mu$ are zero and $v^\alpha_\mu \equiv
v^A_\mu = A^A_\mu$: one recovers the standard results
for the Yang-Mills theory with gauge group $G$.

\section{$\gamma$-cocycles modulo $d$}
\setcounter{equation}{0}

The previous analysis provides the $\gamma$-cohomology
for all values of the ghost numbers.  In particular,
at ghost number zero (no $\omega^{Lie}_I(D)$ in 
(\ref{strict_inv_local})), one sees that the most general
gauge-invariant polynomial is an invariant polynomial in 
the $H$-covariant fields $p^a_{\mu}$, $F^{(v) \alpha}_{\mu \nu}$,
$\tilde{y}^i$ 
and their covariant derivatives.  There are no other strictly
invariant terms.  

To determine the terms that are gauge-invariant only up to
a total divergence, one must, through the descent equation, determine
the possible bottoms (solutions of $\gamma a = 0$) that can be lifted.
Again, this problem has
been systematically studied \cite{DVTV,BDK,DVHTV} with
the result that the most general solution of the
equation $\gamma a + db = 0$ is given, at ghost
number zero, by
\bb
a = c + CS + dm
\ee
where $c$ is stricly invariant ($\gamma c =0$) and thus of the above
form while $CS$ stands for a linear combination with
constant coefficients of the Chern-Simons terms for the
effective gauge group $H$, which
are available only in odd dimensions.  The Chern-Simons terms do not
involve the ``matter" fields $\tilde{y}^i$ or $p^a_\mu$
and their derivatives, but only
the vector potential $v^{\alpha}$ and its field strength.
In particular, one way rewrite the Chern-Simons terms for $G$ as
Chern-Simons terms for $H$ modulo strictly gauge-invariant terms
and total derivatives.  

Similarly, the terms constructed in \cite{deHuRo}, which are defined
in even dimensions,
differ from  strictly
gauge-invariant terms by total divergences.
This was explicitly verified in \cite{HeWi} in the principal
case with abelian gauge group.
A non-Abelian example follows from the analysis of \cite{deLaVP}.
The gauge group is three-dimensional and its Lie algebra reads
\bb
[T_0,T_1]= -2 T_1, \;\; [T_0,T_2]=T_2, \;\; [T_1,T_2]=0.
\ee
The subgroup generated by $T_0$ is non-compact.

The construction of \cite{deHuRo} relies on a non-trivial
cocycle $c_{A,BC}$ of the Lie-algebra cohomology 
with value in the
symmetric tensor product of the adjoint representation,
which we take to be as in \cite{deLaVP}
\bb
c_{1,22}=2, \;\; c_{2,12} = c_{2,21}=-1,
\ee
others vanish.  Besides the Yang-Mills field, we consider the
field $\xi^A$ taking value in the group $G$ (principal case).
The corresponding semi-invariant terms
take the form
\bb
S_{AB}(\xi^A) F^A \wedge F^B + \frac{2}{3}
C_{A,BC}A^A\wedge A^B  \wedge (dA^C - \frac{3}{8}  f^C_{DE} A^D
\wedge A^E)
\label{term}
\ee
One easily verifies that (\ref{term}) is gauge-invariant up to a
non-vanishing surface term provided $S_{AB}= S_{BA}$ transforms as
\bb
\delta_\epsilon S_{AB} = (2 S_{C(A}f^C_{B)D} + C_{D,AB})\epsilon^D
\ee
One may take for $S_{AB}(\xi)$ 
\bb
S_{00} = 2 (\xi^2)^2 \xi^1, \; \;
S_{01} = \frac{1}{2} (\xi^2)^2, \; \; S_{02} = -2 \xi^1 \xi^2,
\ee
\bb
S_{11}= 0, \; \; S_{12} = - \xi^2 , \;\;
S_{22} = 2 \xi^1
\ee
up to terms that transform homogeneously and
clearly lead to strictly gauge-invariant contributions to the action.  
In this case (\ref{term}) reads explicitly
\begin{eqnarray}
& &2(\xi^2)^2 \xi^1 F^0 \wedge F^0 +  (\xi^2)^2 F^0 \wedge F^1
- 4  \xi^1 \xi^2 F^0 \wedge F^2 \nonumber \\
& & -2  \xi^2 F^1 \wedge F^2 + 2 \xi^1 F^2 \wedge F^2
+ 2 A^1 \wedge A^2 \wedge dA^2
\label{notgauge}
\end{eqnarray}

By a direct calculation, one easily checks that
(\ref{notgauge}) differs from the strictly gauge-invariant
term
\bb
2 p^1 \wedge p^2 \wedge dp^2
\label{gaugeinvariant}
\ee
by a total derivative.  Here, the $p^A$'s are
the invariant one-forms
\bb
p^0 = I^0, \; \; p^1 = (\exp{-2 \xi^0}) [I^1 + 2 \xi^1 I^0],
\; \; p^2 = (\exp{\xi^0})[I^2 - \xi^2 I^0]
\ee
with
\begin{eqnarray}
I^0 &=& d \xi^0 - A^0, \nonumber \\
I^1 &=& d \xi^1 - 2 \xi^1 d \xi^0 - A^1 , \nonumber \\
I^2 &=& d \xi^2 + \xi^2 d \xi^0 - A^2.
\end{eqnarray}
Even though $c_{A,BC}$ defines a non-trivial Lie-algebra
cohomological class (for $G$), there is no obstruction in rewriting
(\ref{notgauge}) in a manifestly gauge-invariant way by
adding a total derivative.  The algebraic
obstructions for doing so are given by the Lie-algebra
cohomology of $H$, and are absent here. [In addition,
as mentioned in the conclusions, the De Rham cohomology
of $G$ controls whether the construction done in a star-shaped
neighbourhood of the identity in field space can be extended to
the whole of of the field manifold.  There is no
problem here since the De Rham cohomology of $G$ is trivial;
and indeed, (\ref{gaugeinvariant}) is globally
defined.]

\section{Gauged Wess-Zumino-Witten Terms} 
\setcounter{equation}{0}

The problem investigated in this letter is the problem of finding the
most general term that can be added to the Lagrangian while
preserving $G$-gauge-invariance (up to a total
derivative).  This problem has an indirect
bearing on the problem of gauging the Wess-Zumino-Witten
term, which we first briefly review.

The Wess-Zumino-term \cite{WZ,Witten}
for the ungauged non-linear sigma model in $n$ spacetime dimensions
reads
\bb
W = \int_X \sigma^* h
\label{WZWterm}
\ee
where $X$ is a $(n+1)$-dimensional manifold with spacetime as
boundary and where $h$ is a closed $(n+1)$-form on the
manifold $M$ of the scalar fields that is
invariant under $G$,
\bb
{\cal L}_{\Omega_A} h = 0, \; \; dh =0.
\ee
In (\ref{WZWterm}), $\sigma^* h$ is the pull-back of $h$ to $X$.
One may write $h=db$ locally in field space.  Thus, one
can transform (\ref{WZWterm}) as a $n$-dimensional integral
over spacetime
\bb
W = \int d^nx b_{a_1 \dots a_n}(\xi(x)) \partial_{\alpha_1}
\xi^{a_1} \partial_{\alpha_2} \xi^{a_2} \cdots \partial_{\alpha_n}
\xi^{a_n} \epsilon^{\alpha_1 \dots \alpha_n}
\label{bversion}
\ee
for configurations of the fields in a star-shaped neighbourhood
of $\xi =0$.

Because $h$ is invariant under $G$, the $n$-form $b$
is invariant under $G$ up to a total derivative
\bb
{\cal L}_{\Omega_A} b = d k_A
\ee
This guarantees that the Wess-Zumino-Witten term is (semi-)invariant
under the rigid $G$-transformations generated by the Killing vectors
$\Omega_A^a$.  
Again, it may not be possible to add a total derivative
to the integrand of (\ref{bversion}) to make it strictly $G$-invariant.
There may be algebraic restrictions for doing this, independently
of whether $b$ can be globally defined in the whole of $M$.  For
instance,  a non-zero
translation-invariant (= constant) $(n+1)$-form in $R^k$ is
closed and cannot be written as the exterior derivative of a
translation-invariant $n$-form since the exterior derivative
of a constant $n$-form vanishes.  The obstruction
exists even though the De Rham cohomology of $R^k$
is trivial; it can be detected by a local analysis.

One says that the WZW term (\ref{WZWterm}) can be gauged if one
can add to it terms involving the $G$-connection $A^A_\mu$ so as to
make it invariant under local $G$-transformations up to a
total derivative.  We claim that in even spacetime
dimensions, the WZW term can be gauged if and only if 
one can make the integrand of (\ref{bversion})  {\it strictly} invariant under
$G$-transformations by adding to it a total derivative.  
In that case, the gauging is of course
direct.  If one cannot (locally)
replace $b$ in (\ref{bversion})
by a term that is strictly invariant under
rigid $G$-transformations,
the gauging is impossible.

This is an immediate consequence of our analysis.  Indeed, if the
WZW term can be gauged, then the gauged WZW term must be in the 
list given above
of the terms that can be added to the Lagrangian without destroying the
gauge symmetry.  Thus, the gauged WZW term (if it can be gauged)
must be one of the possible semi-invariant terms for (\ref{gaugetransfsigma}).

In even dimensions, there
is no Chern-Simons term and all semi-invariant terms differ from
strictly invariant ones by  total derivatives.  Thus, in
particular, the gauged 
WZW term (if it exists) takes the form
$WZW_{gauged} = \int (Invariant + dk)$.  
Setting the connection $A^A_\mu$ equal to zero
in this relationship yields the desired result.  Note that the
argument does not work in
odd spacetime dimensions.  There are gaugeable WZW terms that are not described
by invariant forms $b$.  Upon gauging, these lead to the
Chern-Simons terms for $H$, an information that one can use to explicitly
construct them.

Because the conditions for gauging \cite{Hull2,Jacketal}
are not usually formulated in
those terms, we find it instructive to verify explicitly
the (local) equivalence of the gauged WZW term to a strictly $G$-invariant
term in the simplest case, namely, two dimensions.
As stated above, the obstructions to that equivalence can be
detected by a local analysis, so we work in a stratum of the manifold
$M$ of the scalar fields where the
isotropy group is constant.  We consider in fact a region
of the form $B \times
G/H$ where $B$ is a star-shaped open set with $G$-invariant
coordinates $\psi^\Delta$.  We keep explicitly the $B$-factor
to check that it is indeed irrelevant for the present
purposes.

Let $N$ be the counting operator for the $d\psi$'s,
\bb
N = d\psi^\Delta \frac{\partial}{\partial(d\psi^\Delta)}.
\ee
The $3$-form
$h$ splits into a sum
\begin{equation}
h = h^{(0)} + h^{(1)} + h^{(2)} + h^{(3)}
\ee
where $N h^{(k)} = k h^{(k)}$, i.e., $h^{(k)}$ contains $k$ 
$d \psi^\Delta$ and $(3-k)$ $d \xi^a$.  Since the Killing
vectors $\Omega^a_A$ depend only on the $\xi$'s, one
has
$[N, {\cal L}_{\Omega_A}] = 0$.  Furthermore, $d=d_0 + d_1$, $[N,d_0]=
0$, $[N, d_1] = d_1$ with $d_0 = (\partial/\partial \xi^a) d \xi^a$ and
$d_1 = (\partial/\partial \psi^\Delta) d\psi^\Delta$.

Because $dh =0$ and ${\cal L}_{\Omega_A} h =0$, one has
$d_1 h^{(3)} = 0$ and ${\cal L}_{\Omega_A} h^{(3)} = 0$.
The Poincar\'e lemma in $B$ implies $h^{(3)} = d_1 u^{(2)}$ where $u^{(2)}$
is a $2$-form that may be taken to be invariant, ${\cal L}_{\Omega_A}
u^{(2)} =0$.  Indeed, the $\xi^a$ are external parameters for $d_1$
so the standard homotopy for $d_1$ (not $d$ !),
obtained by integrating along rays in $B$, is easily
verified to commute with ${\cal L}_{\Omega_A}$.
If one substracts $du^{(2)}$ from $h$, one gets a $3$-form
$h'$ without components containing three $d \psi$'s, $h=h' 
+ du^{(2)}$, $h' = h^{(0)} + h^{(1)} + h'^{(2)}$.  

By repeating the
analysis for $h'^{(2)}$ and then $h'^{(1)}$, 
one easily concludes that  $h$ takes the
form
\bb
h = \tilde{h} + du
\label{sssss}
\ee
where both $u$ and $\tilde{h}$ are invariant
\bb
{\cal L}_{\Omega_A} u = 0, \; \; \; {\cal L}_{\Omega_A}\tilde{h} = 0
\ee
and where $\tilde{h}$ reduces to its $0$-th component,
\bb
\tilde{h} = \tilde{h}^{(0)}.
\ee
One has also $d\tilde{h} = 0$, which implies both $d_1 \tilde{h} =0$
and $d_0 \tilde{h} = 0$.  The first condition means that
$\tilde{h}$ has no $\psi^\Delta$-dependence and is thus a $3$-form
defined entirely on $G/H$.  Since $h$ is gaugeable if and only if
$\tilde{h}$ is (recall that $u$ in (\ref{sssss}) is $G$-invariant),
we have completely eliminated the $B$-factor.  We can thus assume without
loss of generality that $M$ reduces to a single orbit $G/H$.
This will be done in the sequel; we shall also
drop the tilde on $\tilde{h}$.

As shown in \cite{Hull2,Jacketal},
the $3$-form $h$ on $G/H$ leads to a $G$-gaugeable WZW term if and only if
it fulfills
\bb
i_{\Omega_A}h = dv_A
\label{crucial1}
\ee
for some $1$-forms $v_A$ on $G/H$ that are required to
obey
\bb 
i_{\Omega_A} v_B + i_{\Omega_B} v_A = 0
\label{crucial2}
\ee
as well as
\bb
{\cal L}_{\Omega_A} v_B = f^C_{AB} v_C.
\label{crucial3}
\ee
Now, the algebraic condition (\ref{crucial2}) implies that
$v_A$ takes the form
\bb
v_A = i_{\Omega_A} b
\label{crucial4}
\ee
for some $2$-form $b$ on $G/H$ that is uniquely determined from $v_A$.
Indeed, given $v_A$ and $\Omega_A$,
the equations (\ref{crucial4}), which
read explicitly $\Omega^b_A b_{ab} = v_A^a$, form
a system of linear, inhomogeneous
equations for $b_{ab}$.  The system
possesses at most a solution because the Killing vectors
span the tangent space as $G$ is transitive, so $\Omega^b_A 
c_{ab} = 0$ implies $c_{ab} =0$.  The solution exists because the
consistency conditions of (\ref{crucial4}) are precisely (\ref{crucial2}).

Substituting  (\ref{crucial4}) into (\ref{crucial3}) yields then
\bb
i_{\Omega_B} ({\cal L}_{\Omega_A} b) = 0.
\ee
Again, since the Killing vectors
span the whole tangent space to $G/H$,  this implies
\bb
{\cal L}_{\Omega_A} b = 0. 
\label{crucial5}
\ee
The $2$-form $b$ is thus invariant.  Inserting 
(\ref{crucial4}) into (\ref{crucial1}) and using (\ref{crucial5})
shows then that $H$ is the exterior derivative of an
invariant $2$-form.  A different way to arrive at the same conclusion
is to consider the descent associated with (\ref{bversion})
and to observe that there is no element in $H(\gamma)$
with ghost number two and form-degree zero, or ghost number one
and form-degree one, which could serve as non trivial bottom. Thus,
the descent is effectively trivial and one may redefine
the integrand of (\ref{bversion}) to make it strictly
invariant. [The strictly-invariant expression
equivalent to the gauged WZW term
may however not be easy to work out explicitly in 
practice, or may not be convenient].

\section{Comments and Conclusions} 
\setcounter{equation}{0}

Our results are somewhat disappointing because they show that
the familiar $H$-Chern-Simons terms exhaust all semi-invariant terms
when $H$ is compact or semi-simple.   There are no others.
In particular, in even dimensions, the construction of
\cite{deHuRo} provides semi-invariant terms that turn out to
be equivalent to strictly invariant ones.
In retrospect, this is perhaps not too surprising since one knows that
$G$-invariance boils down to $H$-invariance in the coset space
case $G/H$ with reductive embedding of $H$.  To get new terms, one
needs thus to consider more general situations.

The above analysis assumed that the field variables
of the sigma-model were restricted to a (star-shaped) neighbourhood of
the identity.
One may easily take into account the non-trivial topology
of field space along the lines of \cite{BBH5,HeWi}.
The most convenient way to doing so is to work in the
lifted formulation, in which the sigma-model variables are 
$G$-elements and the right action of $H$ is gauged by means of a 
gauge field without kinetic term.
One just finds, besides the Chern-Simons terms,
additional semi-invariant terms, 
namely the $G$-``winding number"
terms.  These well-known terms are the pull-back
to spacetime of representatives of the
De Rham cohomological classes of $G$.  Like
$\theta$-terms, they can be
written (locally in field space but not globally) as total divergences
and do not contribute therefore to the equations of motion.
Chern-Simons like terms are not of this
type since they do modify the equations of motion and thus
can be detected even in a local approach in field space.  
The details are given in \cite{Wilch}, where the inclusion of the
antifields is also treated and shown not to alter the main conclusions.
One can, also in the presence of antifields,
reduce the problem to the much studied
standard Yang-Mills problem
with gauge group $H$.

\section*{Acknowledgements}

MH is grateful to LPTHE (Orsay) for
warm hospitality 
while this paper was completed.  
This work has been partly supported by the ``Actions de
Recherche Concert{\'e}es" of the ``Direction de la Recherche
Scientifique - Communaut{\'e} Fran{\c c}aise de Belgique", by
IISN - Belgium (convention 4.4505.86) and by
Proyectos FONDECYT 1970151 and 7960001 (Chile).

\end{document}